\documentclass[aps,prb,reprint,superscriptaddress,floatfix,amsmath,amssymb]{revtex4-2}
\usepackage{graphicx}
\usepackage{bm}
\usepackage{booktabs}
\usepackage[colorlinks=true,linkcolor=blue,citecolor=blue,urlcolor=blue]{hyperref}
\usepackage{subfigure}

\newcommand{\Czz}{C_{zz}}

\begin{document}

\title{Hidden Unit Interpretability in RBM Quantum States:\\
Encoding Antiferromagnetic Order in Heisenberg Spin Rings}

\author{Bharadwaj Chowdary Mummaneni}
\email{bharadwaj.chowdary.mummaneni@iao.fraunhofer.de}
\affiliation{Fraunhofer IAO, Nobelstraße 12, 70569 Stuttgart, Germany}

\author{Manas Sajjan}
\email{msajjan@ncsu.edu}
\affiliation{Department of Electrical and Computer Engineering, North Carolina State University, Raleigh, NC 27606, USA}
\affiliation{Quantum Science Center, Oak Ridge National Laboratory, Oak Ridge, TN 37831, USA}

\date{\today}

\begin{abstract}
We investigate how Restricted Boltzmann Machines (RBMs) encode antiferromagnetic order when trained as variational ans\"atze for one-dimensional Heisenberg spin rings with periodic boundary conditions. Through systematic hidden unit analysis and ablation studies on $N=4$ and $N=8$ spin systems, we show that individual hidden units spontaneously specialize to capture staggered magnetization patterns characteristic of antiferromagnetic ground states. Hidden units naturally segregate into two classes: those essential for ground-state energy and correlation structure, and supplementary units providing smaller corrections. Removing important units induces clear energy penalties and disrupts the staggered correlation pattern in $C_{zz}(r)$, whereas removing supplementary units has modest effects. Single-unit analysis confirms that no individual hidden unit reproduces the full antiferromagnetic correlations, indicating that quantum order emerges through collective encoding across the hidden layer. Extending this analysis to $N=8$ through $20$ with hidden unit densities $\alpha = 2$ to $5$ and ten independent seeds per configuration, we find that the fraction of important hidden units decreases with system size, consistent with sublinear growth $m' \sim N^k$ ($k \approx 0.4$). The energy-correlation impact relationship persists for small to moderate system sizes, though it weakens for the largest systems studied. These results provide a quantitative framework for RBM interpretability in quantum many-body systems.

\end{abstract}

\maketitle

\section{Introduction}
Restricted Boltzmann Machines (RBMs) have emerged as powerful variational ans\"atze for quantum many-body systems, achieving high accuracy for ground state calculations across diverse models including the transverse-field Ising model, Heisenberg antiferromagnets, and even topological states \cite{carleo2017solving,deng2017quantum,nomura2017restricted}. Their success stems from the ability to capture volume-law entanglement while remaining computationally tractable through variational Monte Carlo optimization \cite{deng2017quantum,gao2017efficient}. More broadly, neural quantum states have been extended to recurrent architectures \cite{hibat2020recurrent}, autoregressive models \cite{sharir2020deep}, backflow-augmented ans\"atze \cite{luo2019backflow}, and deep network solutions of the electronic Schr\"odinger equation \cite{pfau2020ferminet}. However, understanding how these neural networks internally represent quantum states remains an open challenge \cite{borin2019physical,melko2019restricted}.

The interpretability of neural quantum states has become increasingly important as these methods are applied to frustrated magnets \cite{choo2019frustrated,westerhout2020generalization}, fermionic systems \cite{choo2020fermionic}, and ab initio chemistry \cite{hermann2020deep}. While traditional variational approaches like tensor networks have clear physical interpretations through entanglement structure \cite{orus2014practical,eisert2010area,hastings2007area}, the ``black box'' nature of RBMs obscures the physical mechanisms underlying their representational power. The connection between RBMs and tensor network states \cite{glasser2018string} provides one route to interpretation, and the use of machine learning to classify phases of matter \cite{carrasquilla2017phases,torlai2018tomography} has also spurred interest in understanding what neural networks learn about quantum systems. Systematic studies of individual hidden unit functionality, however, remain scarce.

The one-dimensional Heisenberg antiferromagnet provides an ideal platform for interpretability studies due to its well-characterized ground state properties and the clear physical signature of antiferromagnetic order \cite{bethe1931theorie,haldane1983nonlinear,affleck1989critical,bonner1964linear,sandvik2010computational}. Small spin rings with periodic boundary conditions offer additional advantages: exact diagonalization provides benchmark results \cite{bonner1964linear}, the ground state exhibits robust antiferromagnetic correlations despite finite size effects \cite{haldane1983nonlinear}, and the system size allows detailed analysis of individual hidden unit contributions.

We address the fundamental question: \emph{How do individual hidden units in an RBM encode the antiferromagnetic order of quantum spin systems?} Using systematic ablation studies, a technique widely employed in machine learning interpretability \cite{molnar2020interpretable}, and weight pattern analysis on Heisenberg rings (N=4 and N=8 spins with periodic boundary conditions), we reveal that hidden units spontaneously organize into functionally distinct groups, with a subset learning to represent the staggered magnetization pattern characteristic of antiferromagnetism. We then extend this analysis to larger systems (N=8 to 20) across multiple hidden unit densities $\alpha=2$ to $5$, examining how the fraction of important hidden units and the energy-correlation relationship evolve with system size.

\section{Theoretical Framework}
\subsection{Restricted Boltzmann Machines for Quantum States}
RBMs are generative stochastic neural networks consisting of two layers: visible units representing physical degrees of freedom and hidden units capturing correlations \cite{carleo2017solving,hinton2006fast}. For quantum many-body systems, the RBM serves as a variational wave function ansatz where the amplitude for a spin configuration $\bm{\sigma} = \{\sigma_1, ..., \sigma_N\}$ with $\sigma_i \in \{-1,+1\}$ is given by the marginal distribution over hidden states $\mathbf{h} = \{h_1, ..., h_M\}$ with $h_j \in \{-1,1\}$:

\begin{equation}
\psi_{\text{RBM}}(\bm{\sigma}) = \sum_{\{h_j\}} e^{-E(\bm{\sigma}, \mathbf{h})}
\end{equation}

where the energy function encodes the network architecture:
\begin{equation}
E(\bm{\sigma}, \mathbf{h}) = -\sum_i a_i \sigma_i - \sum_j b_j h_j - \sum_{ij} W_{ji} \sigma_i h_j
\end{equation}

Here $\{a_i\}$ are visible biases, $\{b_j\}$ are hidden biases, and $\{W_{ji}\}$ are visible-hidden coupling weights. After analytically marginalizing over the binary hidden units, the wave function amplitude becomes \cite{carleo2017solving}:

\begin{equation}
\psi(\bm{\sigma}) = e^{\sum_i a_i \sigma_i} \prod_{j=1}^M 2\cosh\left(b_j + \sum_{i=1}^N W_{ji} \sigma_i\right)
\end{equation}

This form is computationally advantageous for variational Monte Carlo sampling, as amplitudes and their derivatives can be evaluated efficiently. The network parameters $\{a_i, b_j, W_{ji}\}$ are optimized through energy minimization using stochastic reconfiguration or similar methods \cite{carleo2017solving,sorella2001green}. 

The hidden unit density $\alpha = M/N$ controls the expressivity of the ansatz. While $\alpha \to \infty$ enables universal approximation \cite{le2008representational}, practical considerations favor moderate densities where $\alpha = 1-4$ provides a balance between accuracy and computational efficiency \cite{borin2019physical}. The parameter $\alpha$ plays a role analogous to the bond dimension in matrix product states, controlling the representational capacity of the ansatz \cite{deng2017quantum}. Recent analysis has shown that RBMs with finite $\alpha$ can capture several orders of perturbation theory in weakly interacting systems \cite{borin2019physical}, explaining their remarkable success on models like the transverse-field Ising chain.

\subsection{Heisenberg Model on a Ring}

We study the spin-1/2 Heisenberg Hamiltonian with periodic boundary conditions:
\begin{equation}
H = J \sum_{i=1}^{N} \vec{S}_i \cdot \vec{S}_{i+1} = J \sum_{i=1}^{N} (S_i^x S_{i+1}^x + S_i^y S_{i+1}^y + S_i^z S_{i+1}^z)
\end{equation}
where $\vec{S}_i$ are spin-1/2 operators, $J=1$ sets the antiferromagnetic coupling strength, and $\vec{S}_{N+1} \equiv \vec{S}_1$ implements the ring geometry. The ground state of this model exhibits staggered magnetization with characteristic $(-1)^i$ alternation in the spin-spin correlation function $C_{zz}(r) = \langle S_i^z S_{i+r}^z \rangle$ \cite{haldane1983nonlinear,affleck1989critical}.

For small even-$N$ rings, the ground state is a spin singlet with exact degeneracy broken only by finite-size effects. The antiferromagnetic correlations decay algebraically with distance as $(-1)^r \ln(r)/r$ but maintain the staggered pattern that serves as a clear signature for RBM analysis \cite{bonner1964linear,johnston1974thermodynamics}. This makes spin rings ideal test systems for probing how neural networks encode magnetic order: the ground state physics is well understood, exact solutions are available for benchmarking, and the system size permits detailed inspection of individual hidden unit contributions.

The connection to RBM representation is particularly relevant because antiferromagnetic correlations require the wave function to exhibit sign structure that alternates with the staggered magnetization pattern \cite{szabo2020sign}. Understanding how RBM hidden units spontaneously learn to encode this alternating structure provides insight into the mechanisms by which neural networks capture quantum magnetic order, with implications for larger and more complex magnetic systems where such detailed analysis becomes computationally prohibitive.

\begin{figure}[t]
\centering
\subfigure[N=4 ring geometry]{%
  \includegraphics[width=0.49\columnwidth]{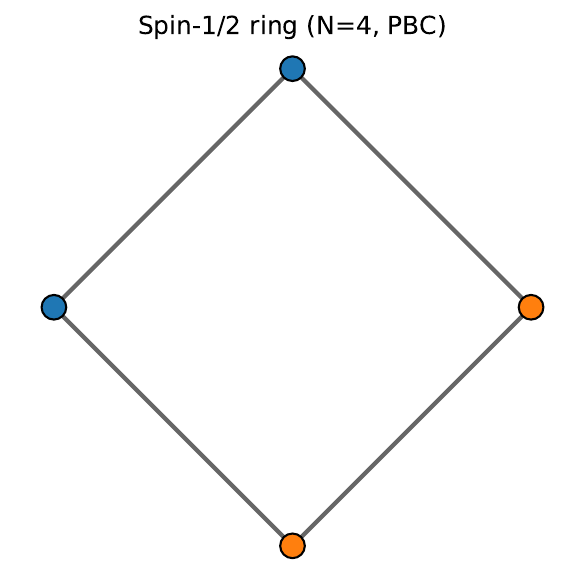}%
  \label{fig:geometry_n4}%
}%
\hfill
\subfigure[N=8 ring geometry]{%
  \includegraphics[width=0.49\columnwidth]{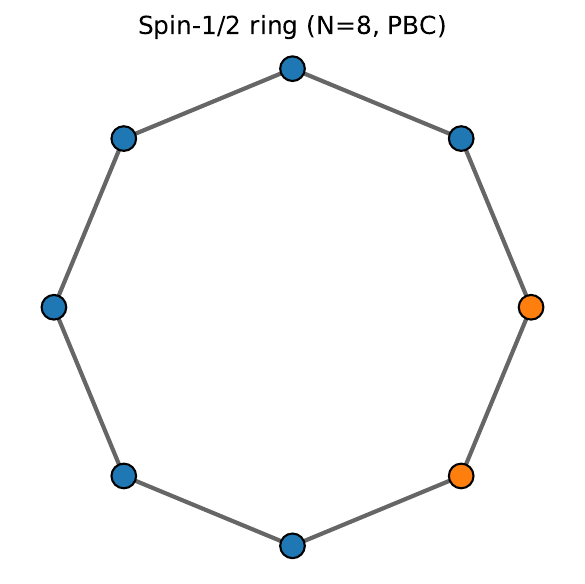}%
  \label{fig:geometry_n8}%
}
\caption{Circular spin systems with periodic boundary conditions. Both small-scale (N=4) and larger-scale (N=8) systems are studied with antiferromagnetic coupling J=1 in the Heisenberg Hamiltonian.}
\label{fig:geometry}
\end{figure}

\subsection{Variational Monte Carlo Optimization}

The RBM parameters are optimized using Variational Monte Carlo (VMC) with stochastic reconfiguration. The energy expectation value:
\begin{equation}
E = \frac{\langle \psi | H | \psi \rangle}{\langle \psi | \psi \rangle} = \sum_{\bm{\sigma}} |\psi(\bm{\sigma})|^2 E_{\text{loc}}(\bm{\sigma})
\end{equation}
is estimated through Monte Carlo sampling, where $E_{\text{loc}}(\bm{\sigma}) = \langle \bm{\sigma} | H | \psi \rangle / \psi(\bm{\sigma})$ is the local energy. The optimization employs Metropolis local sampling (16 chains, 4096 samples/iteration) with the Adam optimizer \cite{kingma2015adam} ($\eta = 10^{-3}$) and stochastic reconfiguration preconditioning \cite{sorella2001green} ($\epsilon_{SR} = 10^{-3}$).

We employ NetKet's implementation of the Heisenberg operator with the Marshall-Peierls sign rule (\texttt{sign\_rule=True}) \cite{marshall1955antiferromagnetism}, which maps the antiferromagnetic ground state to a positive-definite wavefunction in the computational basis. This allows real-valued RBM parameters while preserving the probability distribution over spin configurations characteristic of antiferromagnetic order. Convergence is declared when the relative energy error $\epsilon = |E_{\text{VMC}} - E_{\text{exact}}|/|E_{\text{exact}}|$ falls below a specified threshold; for the systematic study in Sec.~\ref{sec:scaling}, we use $\epsilon < 10^{-3}$ with ten independent seeds per configuration.

\subsection{Analysis Metrics}
To quantify how strongly each hidden unit encodes antiferromagnetic order, we introduce a novel interpretability metric based on the Pearson correlation coefficient \cite{TSERKIS2025130432} between the hidden unit's weight vector and an ideal antiferromagnetic pattern. The AFM pattern score for hidden unit $j$ is defined as:
\begin{align*}
S_{AFM}^{(j)} &= r(\mathbf{W}_j, \mathbf{s}_{AFM}) \\
\\
&= \frac{\sum_{i=1}^N (W_{ji} - \bar{W}_j)(s_i - \bar{s})}{\sqrt{\sum_{i=1}^N (W_{ji} - \bar{W}_j)^2} \sqrt{\sum_{i=1}^N (s_i - \bar{s})^2}} \label{eq:afm_score}
\end{align*}

where the terms are defined as follows:
\begin{itemize}
\item $\mathbf{W}_j = \{W_{j1}, W_{j2}, ..., W_{jN}\}$ is the weight vector connecting hidden unit $j$ to all $N$ visible spins
\item $W_{ji}$ is the connection weight from visible spin $i$ to hidden unit $j$
\item $\bar{W}_j = \frac{1}{N}\sum_{i=1}^N W_{ji}$ is the mean weight for hidden unit $j$
\item $\mathbf{s}_{AFM} = \{(-1)^1, (-1)^2, ..., (-1)^N\}$ is the ideal staggered magnetization pattern
\item $s_i = (-1)^i$ is the expected spin orientation at site $i$ for perfect antiferromagnetic order
\item $\bar{s} = \frac{1}{N}\sum_{i=1}^N s_i$ is the mean of the reference AFM pattern (which equals zero for even $N$)
\item The numerator $\sum_{i=1}^N (W_{ji} - \bar{W}_j)(s_i - \bar{s})$ measures the covariance between weights and the AFM pattern
\end{itemize}

This correlation coefficient ranges from -1 to +1, where we empirically classify:
\begin{itemize}
\item $|S_{AFM}^{(j)}| \approx 1$: Strong antiferromagnetic encoding (positive or negative phase)
\item $|S_{AFM}^{(j)}| > 0.6$: Significant AFM pattern correlation  
\item $|S_{AFM}^{(j)}| \approx 0$: No AFM encoding (uniform, random, or other patterns)
\end{itemize}

In practice, we compute this using NumPy's \texttt{corrcoef} function, which implements the Pearson correlation coefficient efficiently \cite{harris2020numpy}. Units with high $|S_{AFM}^{(j)}|$ scores are identified as primarily encoding the antiferromagnetic order parameter.

\textbf{Hidden unit ablation} quantifies the impact of removing unit $j$:
\begin{equation}
\Delta E^{(j)} = E_{\text{removed}} - E_{\text{full}}
\end{equation}
\begin{equation}
\Delta \Czz^{(j)}(r) = \Czz^{\text{removed}}(r) - \Czz^{\text{full}}(r)
\end{equation}

\section{Results and Analysis}

\subsection{Energy Convergence and Ground State Quality}

\begin{figure*}[t]
\centering
\subfigure[N=4 spin system ]{%
  \includegraphics[width=0.5\textwidth]{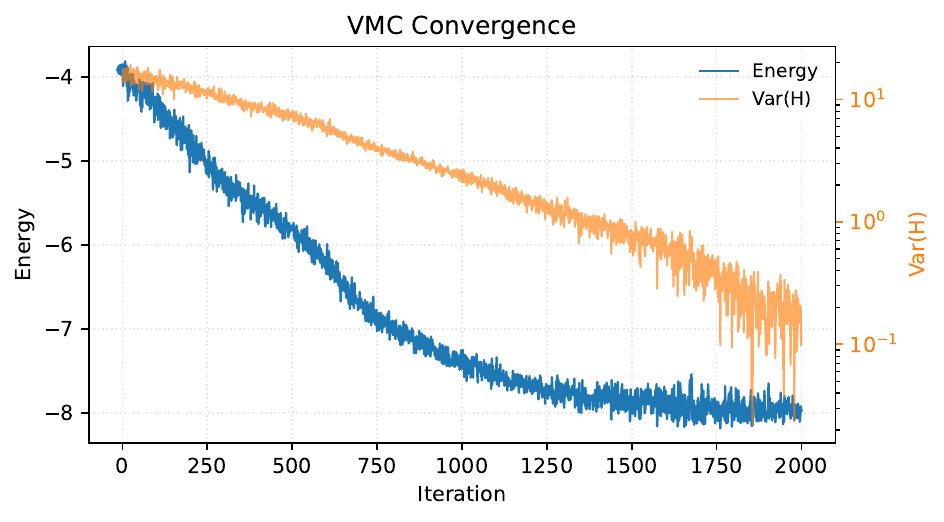}%
  \label{fig:convergence_n4}%
}%
\hfill
\subfigure[N=8 spin system ]{%
  \includegraphics[width=0.5\textwidth]{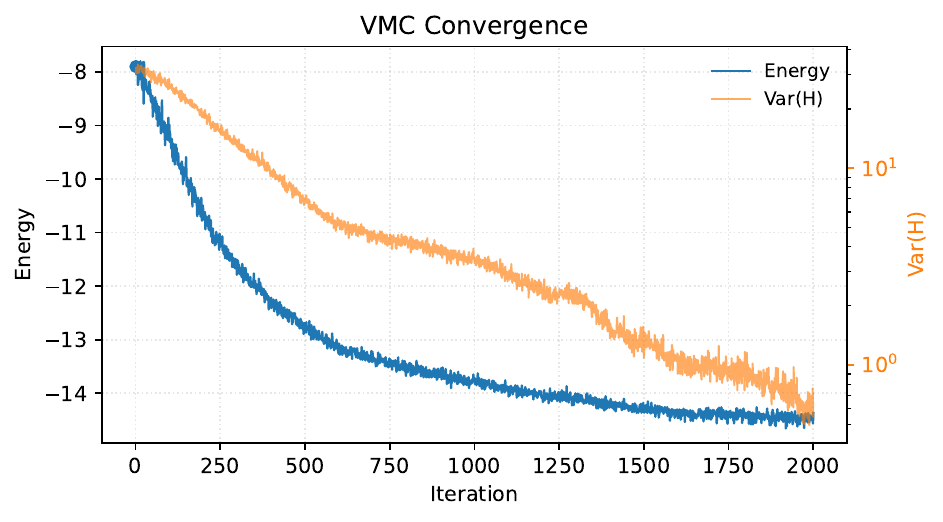}%
  \label{fig:convergence_n8}%
}
\caption{Energy convergence during VMC optimization demonstrating how the RBM ansatz successfully minimizes the Heisenberg Hamiltonian. Both systems show smooth convergence to nearly analytical ground state values: N=4 reaching $E/N=-1.061$ and N=8 reaching $E/N=-1.101$. The variance reaches $10^{-1}$ for the smaller system and remains at $10^{0}$ for the larger system, confirming high-quality variational states.}
\label{fig:convergence}
\end{figure*}

The RBM ansatz demonstrates remarkable efficiency in optimizing the spin Hamiltonian (Fig.~\ref{fig:convergence}). Both systems achieve smooth convergence to energies within 1\% of analytical ground state values. The variance evolution reveals an interesting size-dependent behavior: the smaller N=4 system achieves variance of order $10^{-1}$, while the larger N=8 system stabilizes at $10^{0}$, reflecting the increased complexity of representing larger entangled states.

\subsection{Emergence of Antiferromagnetic Correlations}

\begin{figure*}[t]
\centering
\subfigure[N=4: evolution of $\Czz(r)$]{%
  \includegraphics[width=0.49\textwidth]{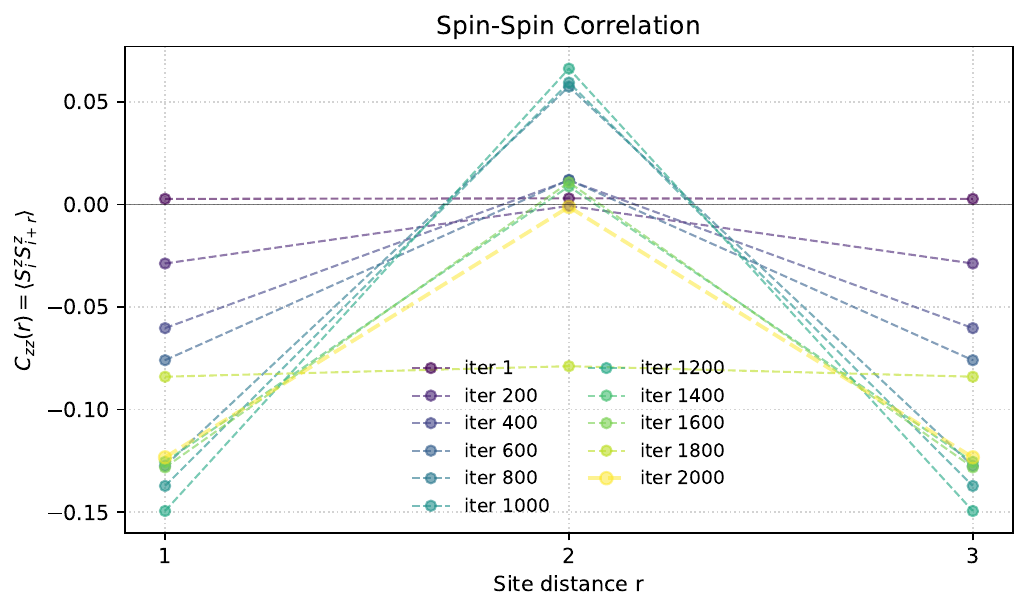}%
  \label{fig:correlations_n4}%
}%
\hfill
\subfigure[N=8: evolution of $\Czz(r)$]{%
  \includegraphics[width=0.49\textwidth]{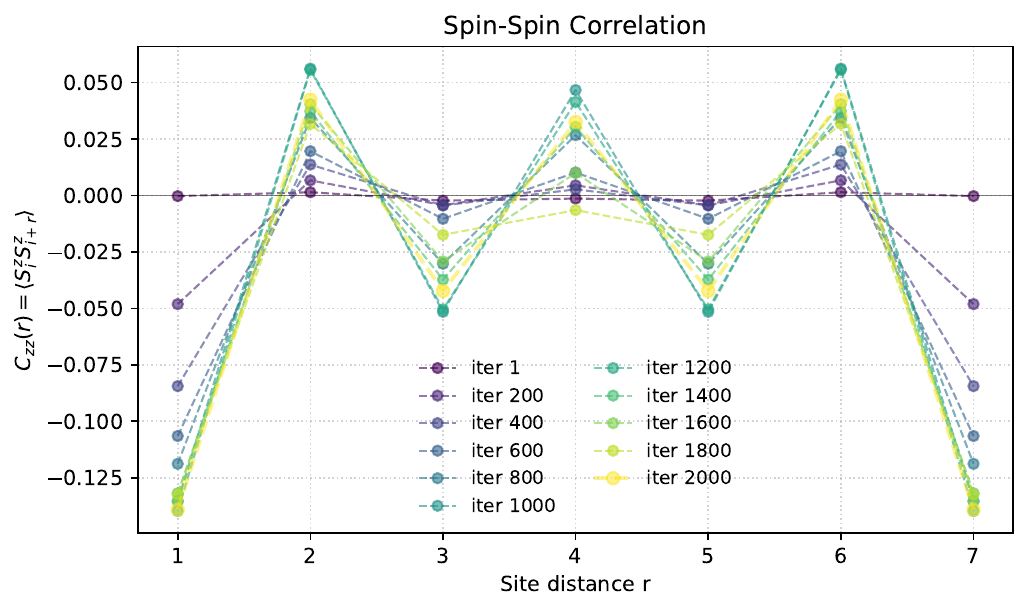}%
  \label{fig:correlations_n8}%
}
\caption{Evolution of spin-spin correlations during training. As iterations progress, the antiferromagnetic nature emerges, with clear $(-1)^r$ oscillations established by convergence. This illustrates how the RBM network progressively learns the correlational structure of the quantum ground state.}
\label{fig:correlations}
\end{figure*}

The correlation function $\Czz(r) = \frac{1}{N}\sum_i \langle S_i^z S_{i+r}^z \rangle$ evolves from random values to clear antiferromagnetic patterns (Fig.~\ref{fig:correlations}). As training progresses, the characteristic $(-1)^r$ oscillations emerge naturally, demonstrating how the RBM network discovers and encodes the system's correlational structure without explicit guidance. This emergent understanding of quantum correlations, while perhaps expected theoretically, demonstrates how neural networks can learn complex quantum phenomena from energy minimization alone.

\subsection{Hidden Unit Specialization and AFM Pattern Encoding}

\begin{figure*}[t]
\includegraphics[width=0.7\textwidth]{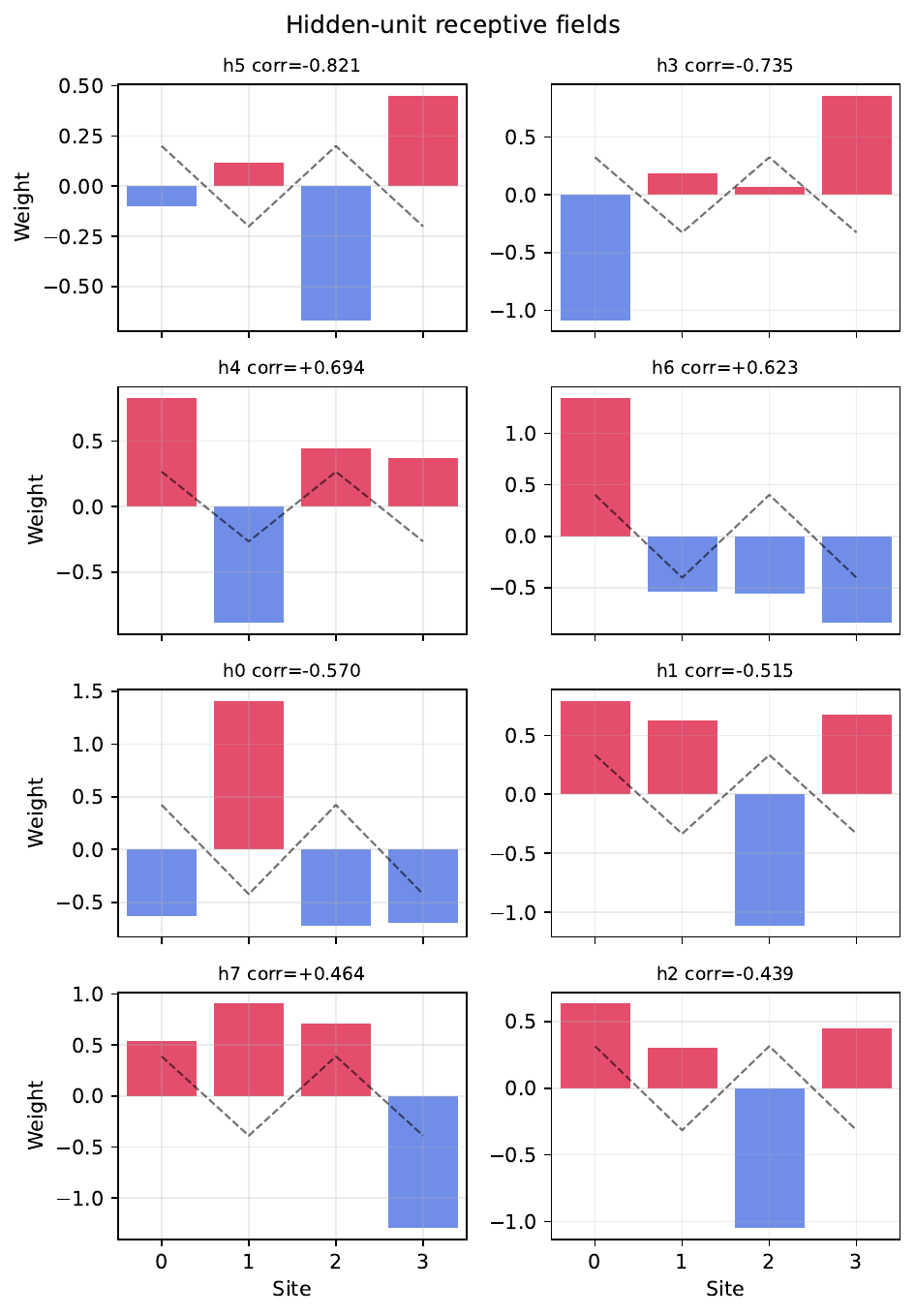}
\caption{Representative hidden unit weight patterns for the N=4 system with 8 hidden units. Units 3 and 5 exhibit clear antiferromagnetic behavior with oscillating patterns and absence of strong ferromagnetic order. Unit 3 shows higher magnitude weights than unit 5, explaining its stronger impact on both energy and correlations. The worst-performing units share a common feature: three sites with ferromagnetic alignment.}
\label{fig:hidden_patterns}
\end{figure*}

Application of the Pearson correlation analysis reveals how RBMs organize their hidden units to encode quantum many-body states. For the N=4 system with 8 hidden units, we find that several units develop strong correlations with the antiferromagnetic pattern, indicating specialized encoding of the dominant ground-state order parameter.

The neural network distributes the computational burden across its hidden units, with some specializing in AFM pattern encoding while others handle long-range correlations and normalization tasks. This emergent division of labor demonstrates the network's ability to automatically discover an efficient representation strategy without explicit guidance.

Detailed analysis of the N=4 system reveals the physical significance of these correlation scores (Fig.~\ref{fig:hidden_patterns}). Hidden units 3 and 5, with the highest $S_{AFM}$ scores, display clear oscillating weight patterns alternating between positive and negative values across lattice sites—the hallmark of antiferromagnetic encoding. Unit 3's correlation score of $S_{AFM}^{(3)} = 0.735$, combined with its larger weight magnitudes, likely contributes to its dominant impact on both energy and correlations. In contrast, units with low correlation scores consistently exhibit patterns such as ferromagnetic alignment across three consecutive sites, demonstrating how the correlation metric successfully identifies units that fail to capture the essential physics.

\subsection{Energy Impact of Hidden Unit Ablation}

\begin{figure*}[t]
\centering
\subfigure[N=4: energy change per hidden unit]{%
  \includegraphics[width=0.49\textwidth]{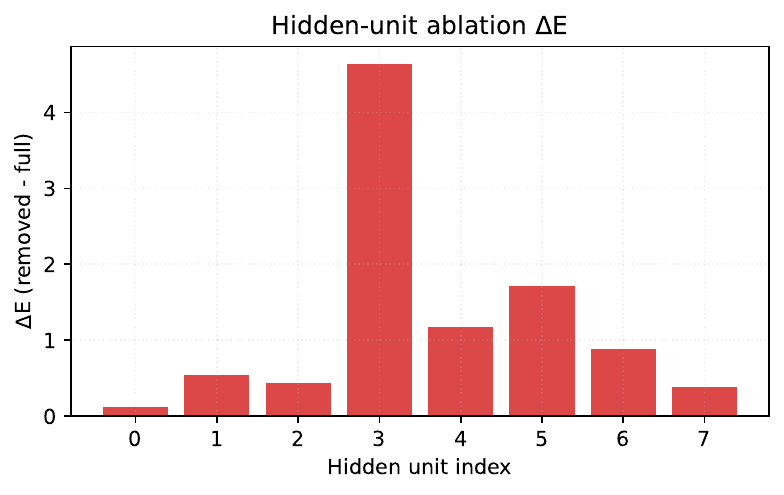}%
  \label{fig:ablation_energy_n4}%
}%
\hfill
\subfigure[N=8: energy change per hidden unit]{%
  \includegraphics[width=0.49\textwidth]{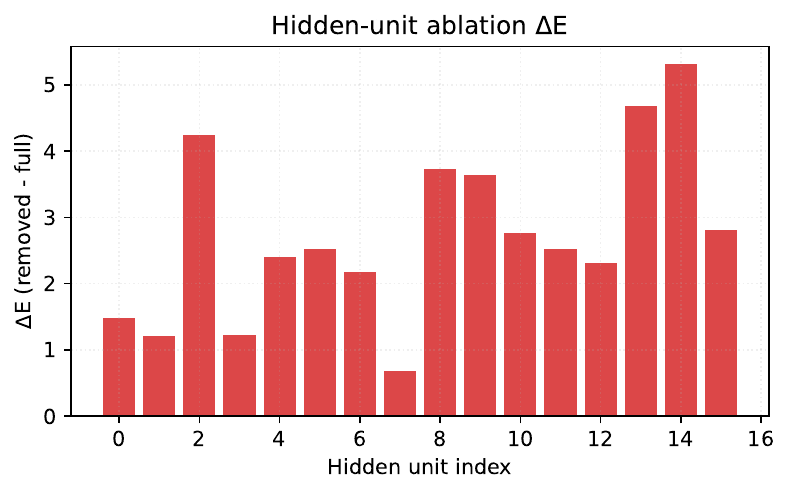}%
  \label{fig:ablation_energy_n8}%
}
\caption{Energy change upon removing individual hidden units. In the N=4 system, unit 3 is critically important, causing >50\% error when removed. Unit 5 also impacts energy significantly, while unit 0 has minimal effect. The N=8 system shows distributed importance across multiple units, with units 14 and 13 having the highest impact.}
\label{fig:ablation_energy}
\end{figure*}

The ablation analysis reveals striking differences between system sizes (Fig.~\ref{fig:ablation_energy}). In the N=4 system, hidden unit 3 emerges as critically important and its removal causes more than 50\% error in energy calculation. Unit 5 also significantly impacts energy, while unit 0 has negligible effect. 

The N=8 system exhibits a more distributed importance pattern, with multiple hidden units contributing significantly to energy accuracy. Units 14 and 13 lead in importance, but unlike the N=4 case, no single unit dominates completely. This suggests that larger systems require more collective encoding of quantum properties.

\subsection{Correlation Pattern Disruption}

\begin{figure*}[t]
\centering
\subfigure[N=4 comparison]{
  \includegraphics[width=\textwidth]{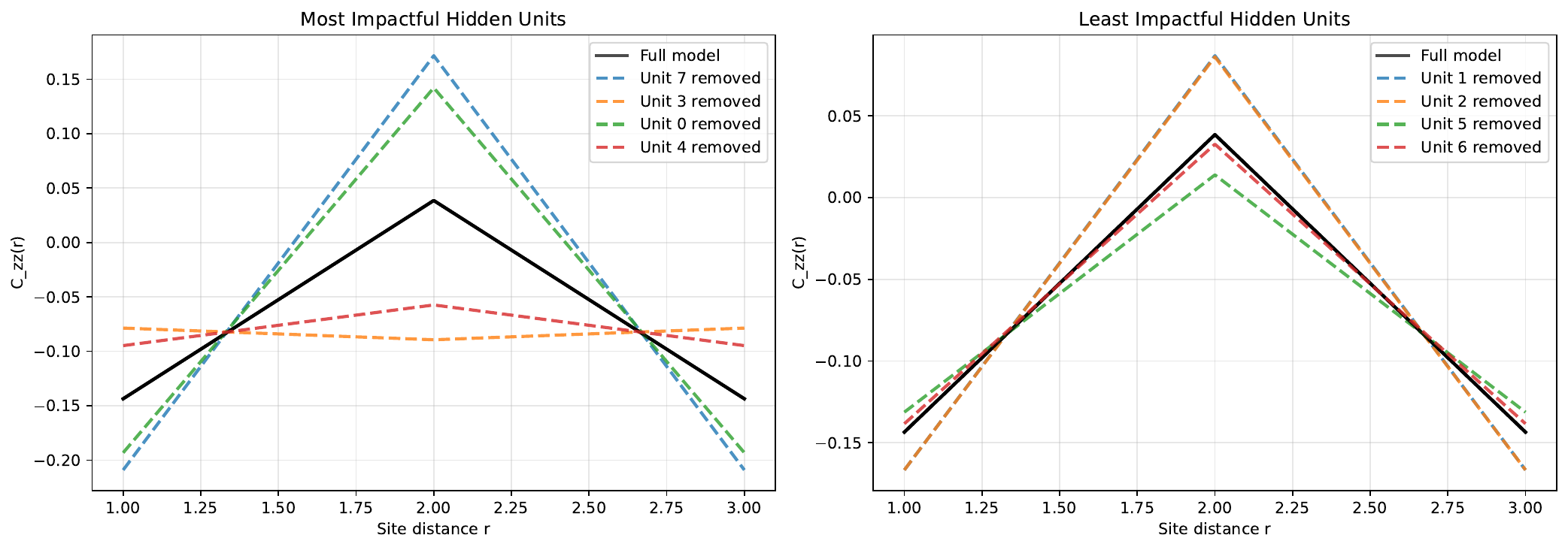}%
  \label{fig:ablation_corr_4}%
}%
\hfill
\subfigure[N=8 comparison]{
  \includegraphics[width=\textwidth]{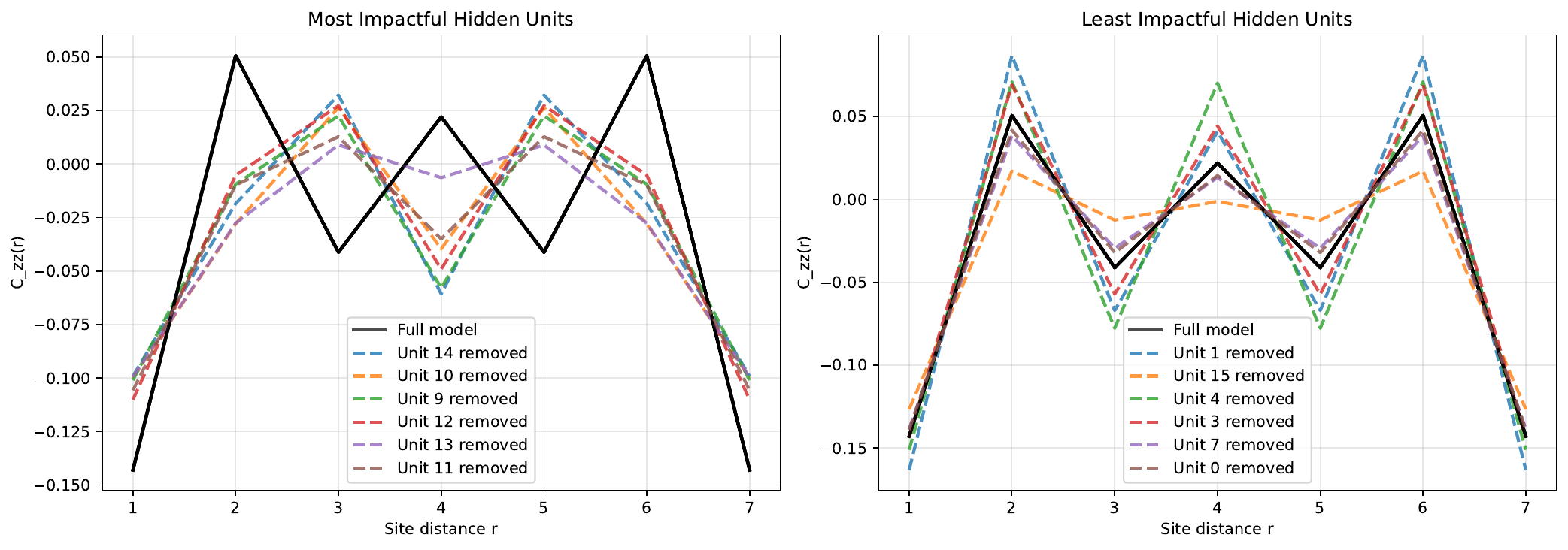}%
  \label{fig:ablation_corr_8}%
}
\caption{Changes in correlation functions upon hidden unit removal. For N=4, removing unit 3 completely destroys the AFM pattern—the system loses antiferromagnetic character entirely. Other important units affect correlation magnitudes while preserving the pattern. For N=8, impactful units show phase-inverted patterns (180° phase shift), while less impactful units preserve the AFM pattern with slight magnitude perturbations.}
\label{fig:ablation_correlations}
\end{figure*}

The correlation analysis reveals profound insights into how hidden units encode quantum order (Fig.~\ref{fig:ablation_correlations}). In the N=4 system, removing unit 3 has significant effects as the entire antiferromagnetic pattern vanishes, not just its magnitude. This unit is essential for maintaining quantum order. Other important units affect correlation amplitudes while preserving the oscillatory pattern. Less impactful units barely disturb the correlations.

The N=8 system exhibits an intriguing phenomenon: removing impactful units produces correlations with opposite phase (180° shift), showing spin-down where the original has spin-up at middle sites. This suggests these units encode specific phase information. Less impactful units preserve the AFM pattern while slightly perturbing magnitudes, indicating they provide fine-tuning rather than fundamental structure.

\subsection{Single Unit Analysis: Collective Quantum Encoding}

\begin{figure*}[t]
\centering
\subfigure[N=4 single unit analysis]{
  \includegraphics[width=\textwidth]{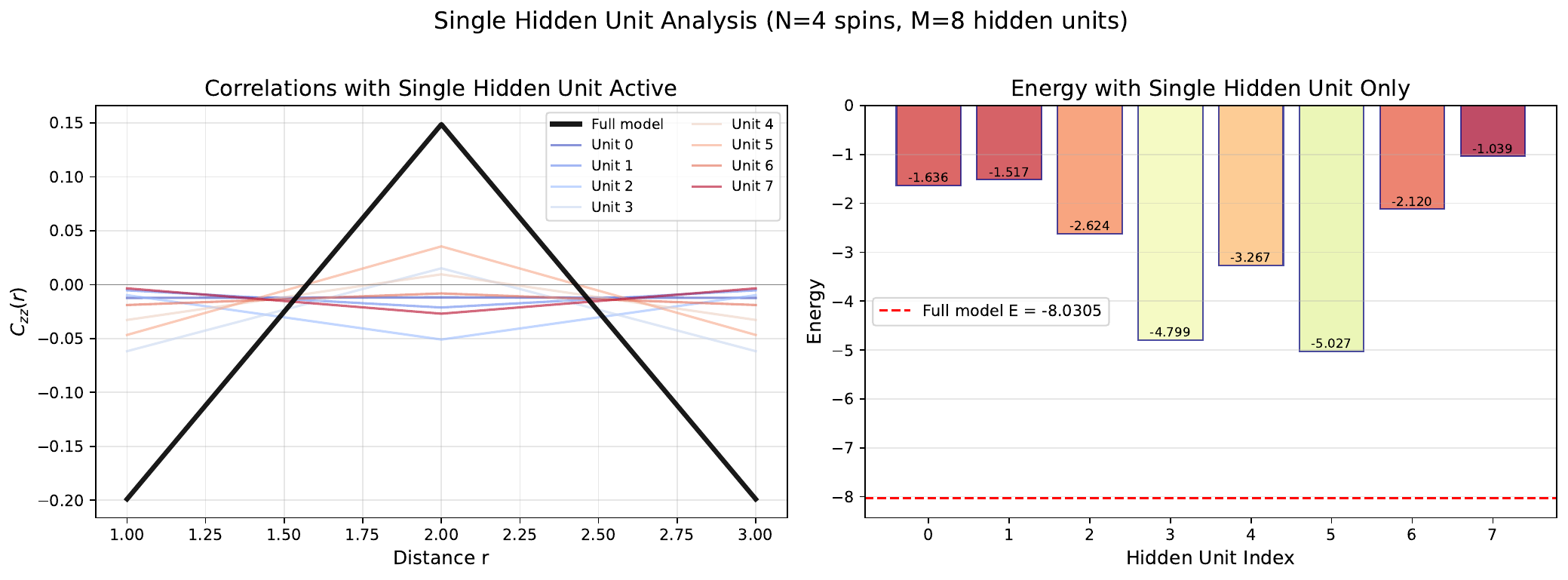}%
  \label{fig:single_unit_n4}%
}%
\hfill
\subfigure[N=8 single unit analysis]{
  \includegraphics[width=\textwidth]{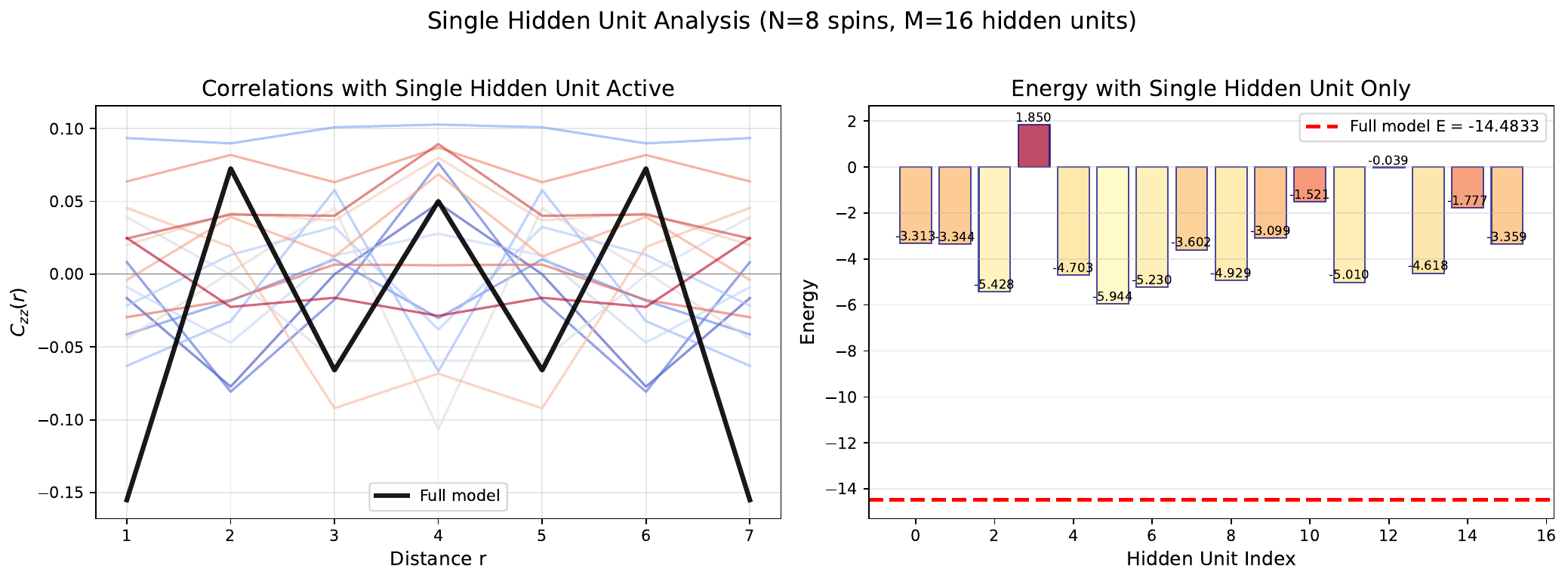}%
  \label{fig:single_unit_n8}%
}
\caption{Single hidden unit analysis reveals collective encoding necessity. For N=4, units 3 and 5 have the largest energy impact when isolated but cannot recover AFM correlations at all. For N=8, no single unit produces significant impact on either energy or correlations. This demonstrates that hidden units must work together in a superposition-like manner to encode quantum properties.}
\label{fig:single_unit}
\end{figure*}

The single unit retention analysis provides definitive evidence for collective quantum encoding (Fig.~\ref{fig:single_unit}). In the N=4 system, even the most important units (3 and 5) fail completely to reproduce antiferromagnetic correlations when isolated, despite their significant energy contributions. For the N=8 system, no individual unit can produce meaningful impact on either energy or correlations alone.

This reveals a fundamental principle: hidden units encode quantum states through superposition-like collective action. The antiferromagnetic order emerges only from the interplay of multiple specialized units, analogous to how quantum states require superposition of basis states. No single classical unit can capture the quantum correlations indicating that they must work in unison.

\subsection{Universal Energy-Correlation Relationship}

\begin{figure*}[t]
\centering
\subfigure[N=4 energy-correlation relationship]{%
  \includegraphics[width=0.5\textwidth]{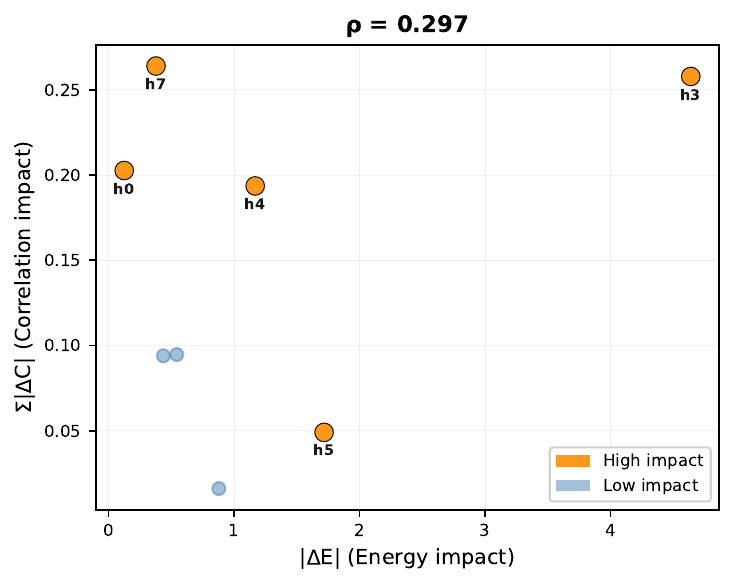}%
  \label{fig:universal_n4}%
}%
\hfill
\subfigure[N=8 energy-correlation relationship]{%
  \includegraphics[width=0.5\textwidth]{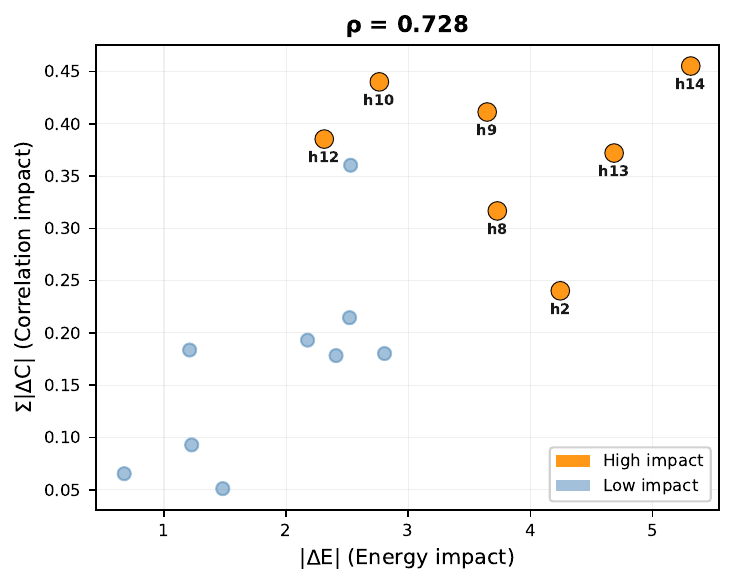}%
  \label{fig:universal_n8}%
}
\caption{Universal relationship between energy and correlation impact. Units positioned in the top-right are most impactful for both properties. N=4: units 3, 5, 7, 0, 4 show high impact. N=8: units 14, 13, 10, 12, 8, 2 are most important. Unit 3 (N=4) and unit 14 (N=8) emerge as singular most important units for their respective systems.}
\label{fig:universal}
\end{figure*}

The combined analysis of energy and correlation impacts reveals which hidden units are globally important (Fig.~\ref{fig:universal}). Units positioned far in the top-right corner impact both energy and correlation spectra most significantly. 

For the N=4 system, units 3, 5, 7, 0, and 4 emerge as impactful, with unit 3 standing alone as the most critical. In the N=8 system, units 14, 13, 10, 12, 8, and 2 show high importance, with unit 14 leading. The strong correlation  between energy and correlation impacts suggests a fundamental connection: units that maintain ground state energy also preserve physical correlations.

\section{Systematic Study Across System Sizes}
\label{sec:scaling}

To test whether the interpretability patterns observed for N=4 and N=8 persist in larger systems, we study even spin rings from N=8 to N=20 with hidden unit densities $\alpha \in \{2, 3, 4, 5\}$. For each configuration, ten independent training runs with different random seeds yield 280 total models. We use a convergence threshold $\epsilon = |E_{\text{VMC}} - E_{\text{exact}}|/|E_{\text{exact}}| < 10^{-3}$; 279 of 280 models converge within this threshold, giving near-complete coverage across all system sizes. All analysis below uses this converged set.

We classify hidden unit $j$ as \emph{important} (belonging to the set $m'$) if removing it causes either a fractional energy change $|\Delta E^{(j)}|/|E_{\text{full}}| > \tau$ or a total correlation disruption $\sum_r |\Delta C_{zz}^{(j)}(r)| / \sum_r |C_{zz}^{\text{full}}(r)| > \tau$, with threshold $\tau = 0.3$. The remaining $M - m'$ units are classified as unimportant. This threshold is not derived from first principles; rather, it is chosen to lie in the regime where the classification meaningfully separates units (at $\tau \lesssim 0.1$, nearly all units are classified as important) while retaining enough signal for scaling analysis (at $\tau \gtrsim 0.5$, most units are classified as unimportant and the power-law fits degrade). The sensitivity of our results to this choice is examined in Fig.~\ref{fig:threshold_sensitivity}.

\subsection{Energy Recovery and Importance Classification}

\begin{figure}[t]
\centering
\includegraphics[width=\columnwidth]{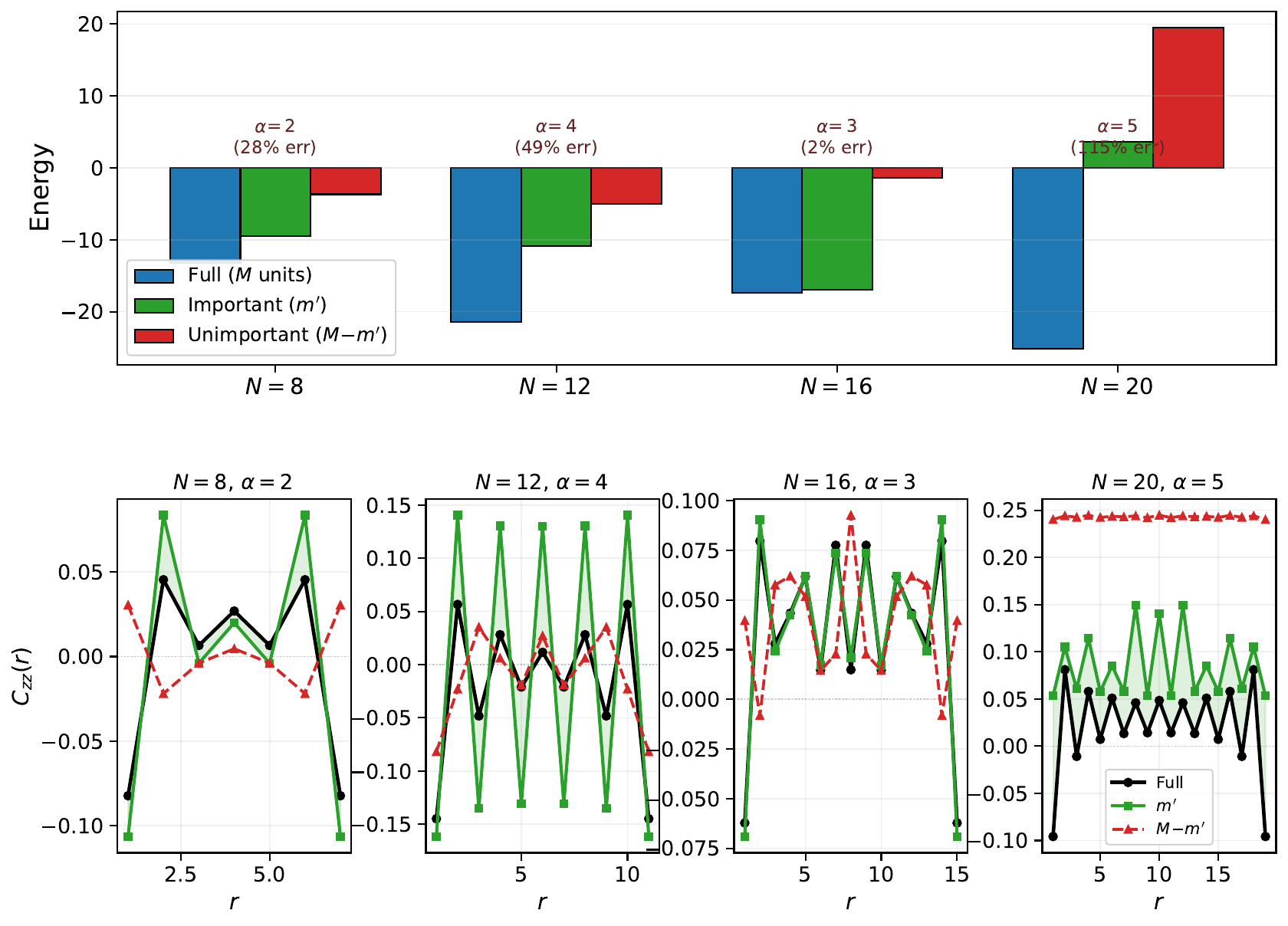}
\caption{Energy decomposition into contributions from the $m'$ important and $M - m'$ unimportant hidden units, for system sizes $N=8$, $12$, $16$, and $20$. For each $N$, the hidden unit density $\alpha$ yielding the closest energy recovery is selected. The bottom panels show the spin-spin correlation $C_{zz}(r)$ for each group, with the shaded region indicating the gap between the full model and the important-unit subset.}
\label{fig:energy_recovery}
\end{figure}

Figure~\ref{fig:energy_recovery} shows the energy recovered by retaining only the $m'$ important units versus the full model. The quality of recovery varies considerably across system sizes: at $N=16$ ($\alpha=3$), the $m' = 40$ important units recover the energy to within 2.4\%, while at $N=8$, $12$, and $20$ the recovery errors are substantially larger (28\%, 49\%, and 114\% respectively). The large errors at $N=12$ and $N=20$ indicate that the retained subset, while capturing the units with the largest individual ablation impacts, does not account for collective interactions among units that contribute to the full energy. Nevertheless, the correlation panels show that the important units consistently reproduce the oscillatory structure of $C_{zz}(r)$, while the unimportant units yield flat or inverted correlation profiles. This confirms that the importance classification captures physically meaningful structure even when the energy recovery is imperfect.



\subsection{Scaling of Important Hidden Units}

\begin{figure*}[t]
\centering
\subfigure[Fraction $m'/M$ vs system size $N$]{%
  \includegraphics[width=0.33\textwidth]{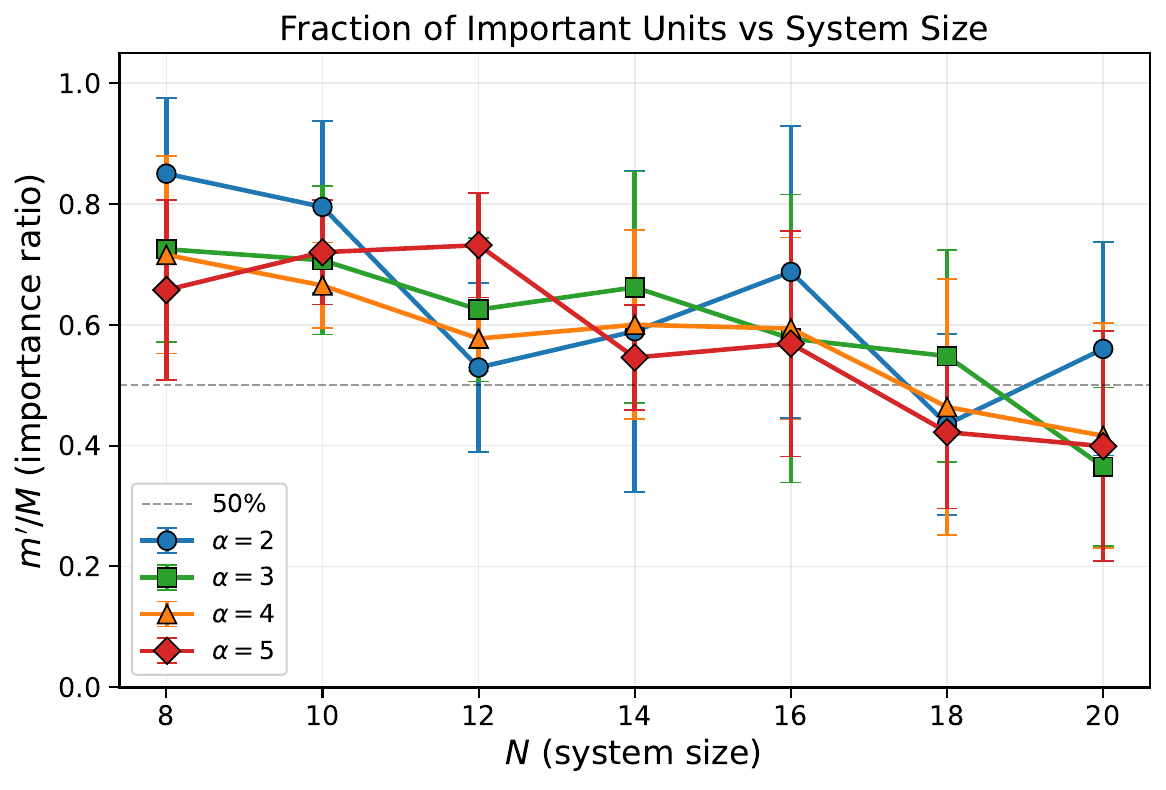}%
  \label{fig:ratio_vs_N}%
}%
\hfill
\subfigure[Number of important units $m'$ vs $N$]{%
  \includegraphics[width=0.33\textwidth]{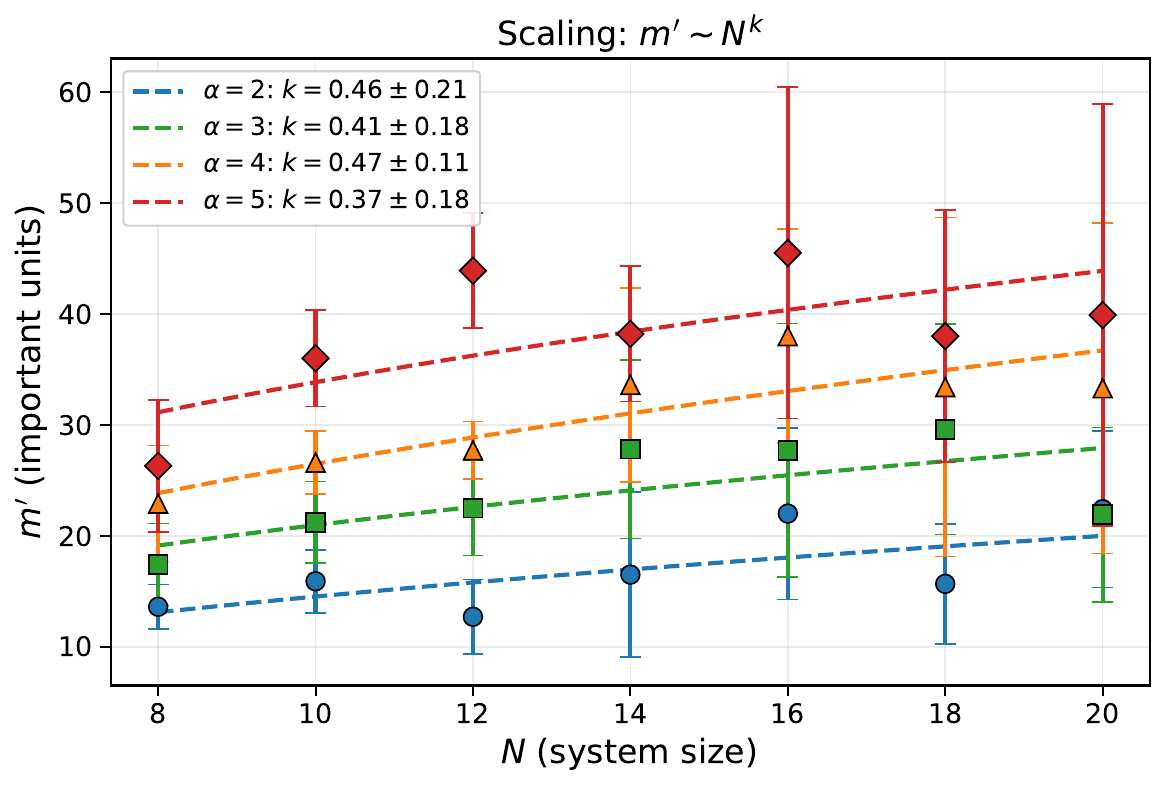}%
  \label{fig:m_prime_vs_N}%
}%
\hfill
\subfigure[Power-law scaling exponent $k$]{%
  \includegraphics[width=0.33\textwidth]{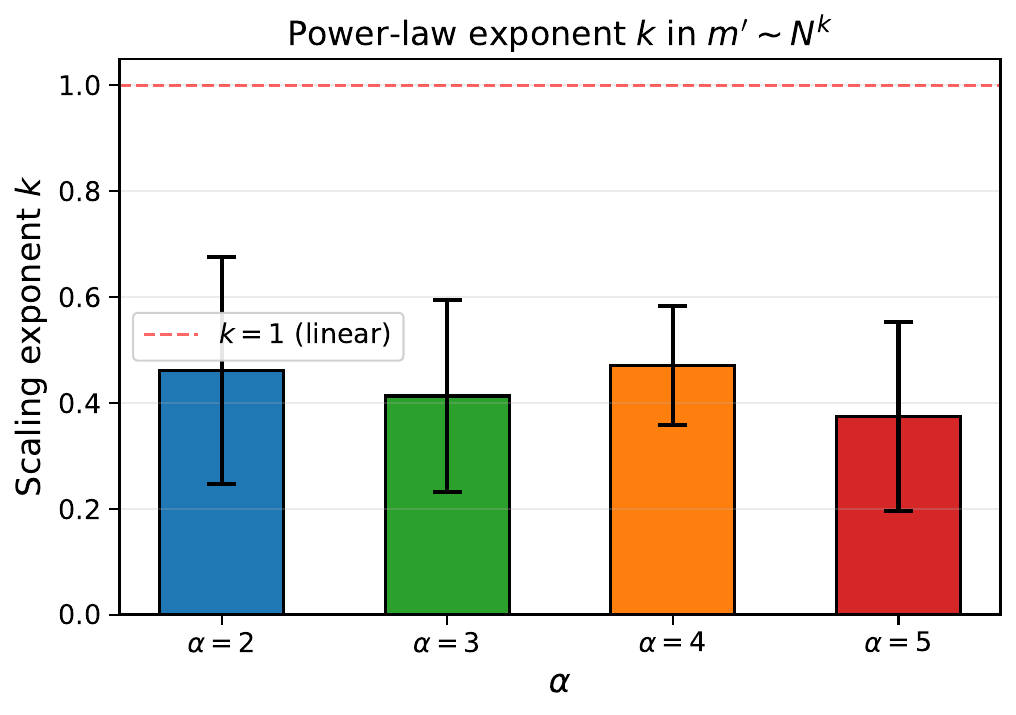}%
  \label{fig:scaling_exponent}%
}
\caption{Scaling of important hidden units with system size (10 seeds per configuration). (a)~The fraction $m'/M$ of important units shows a general decrease with $N$ for all $\alpha$ values. (b)~Power-law fits $m' \sim N^k$ (dashed lines) to the seed-averaged data. (c)~Fitted exponents $k$ with standard errors; all values lie below unity, consistent with sublinear growth. The fit quality varies from $R^2 = 0.47$ ($\alpha = 5$) to $R^2 = 0.78$ ($\alpha = 4$).}
\label{fig:scaling}
\end{figure*}

A natural question is how the number of important hidden units $m'$ scales with system size. Figure~\ref{fig:scaling} presents this analysis across all four hidden unit densities, with ten seeds per configuration providing improved statistics over our preliminary five-seed study. The fraction $m'/M$ shows a clear downward trend with increasing $N$ [Fig.~\ref{fig:ratio_vs_N}], though seed-to-seed variability remains substantial.

Fitting $m' = a \cdot N^k$ to the seed-averaged data [Fig.~\ref{fig:m_prime_vs_N}] yields exponents $k = 0.46 \pm 0.21$ ($\alpha = 2$), $0.41 \pm 0.18$ ($\alpha = 3$), $0.47 \pm 0.11$ ($\alpha = 4$), and $0.37 \pm 0.18$ ($\alpha = 5$). The exponents are consistent across all four $\alpha$ values and all lie well below unity. The fit quality is best for $\alpha = 4$ ($R^2 = 0.78$) and moderate for the others ($R^2 = 0.47$ to $0.51$). Since $M = \alpha N$ grows linearly, the decreasing fraction $m'/M$ implies that not all additional hidden units are equally utilized as the system grows.

\begin{figure}[t]
\centering
\includegraphics[width=\columnwidth]{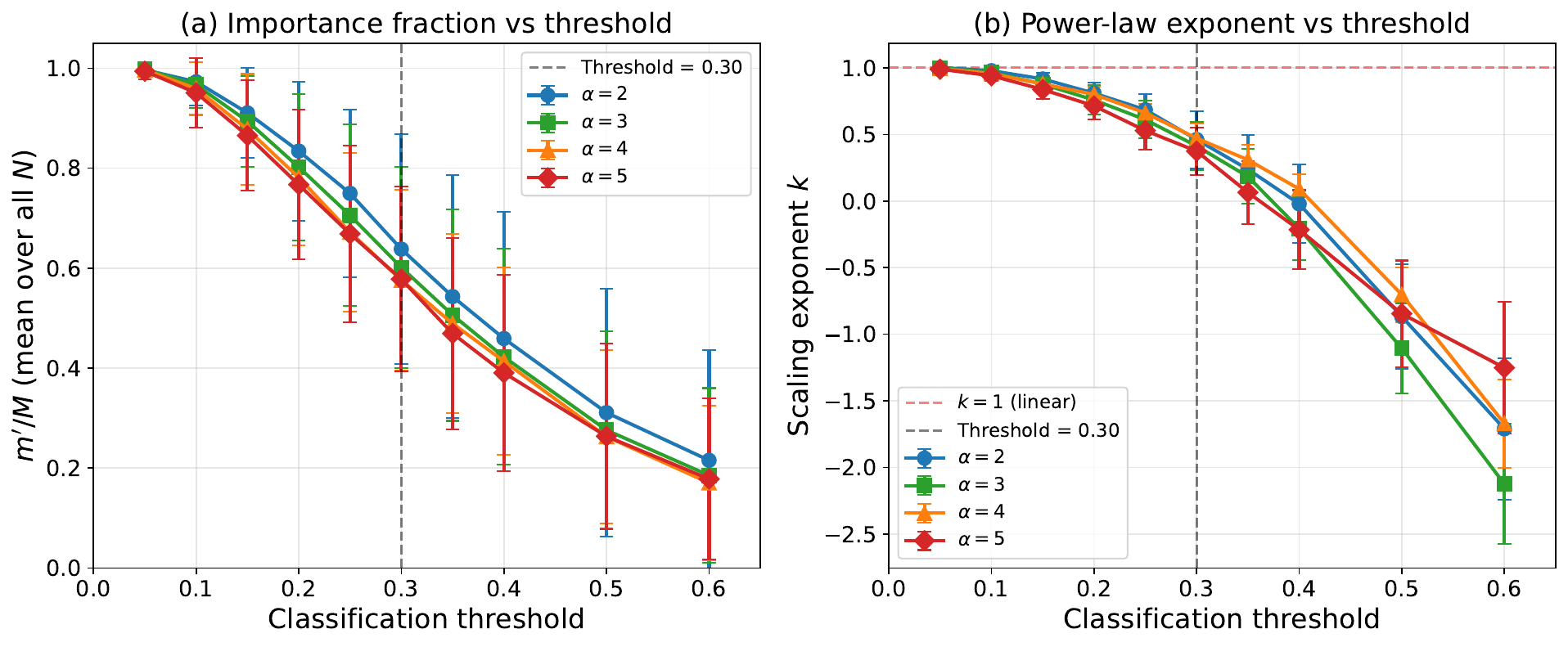}
\caption{Sensitivity of the importance classification to the threshold $\tau$. (a)~The mean fraction $m'/M$ of important units (averaged over all system sizes) decreases smoothly with $\tau$ for all $\alpha$ values; the dashed line marks the threshold $\tau = 0.3$ used throughout this work. (b)~The power-law exponent $k$ extracted from $m' \sim N^k$ fits. Sublinear scaling ($k < 1$) is observed for $\tau \gtrsim 0.15$ across all $\alpha$ values, demonstrating that the qualitative conclusion is not an artifact of the particular threshold chosen.}
\label{fig:threshold_sensitivity}
\end{figure}

To verify that our conclusions do not depend sensitively on the threshold $\tau = 0.3$, we repeat the classification and power-law fits across a range of thresholds from 0.05 to 0.60 (Fig.~\ref{fig:threshold_sensitivity}). The fraction $m'/M$ varies smoothly with $\tau$, and the transition from $k \approx 1$ (all units important) to $k < 1$ (sublinear scaling) occurs around $\tau \approx 0.15$ for all $\alpha$ values. The sublinear regime persists through $\tau \approx 0.35$, beyond which the fits become unreliable as too few units remain classified as important. The value $\tau = 0.3$ lies well within this stable sublinear regime.

\subsection{Dependence on Hidden Unit Density $\alpha$}

\begin{figure}[t]
\centering
\subfigure[Energy convergence across $\alpha$]{%
  \includegraphics[width=0.49\textwidth]{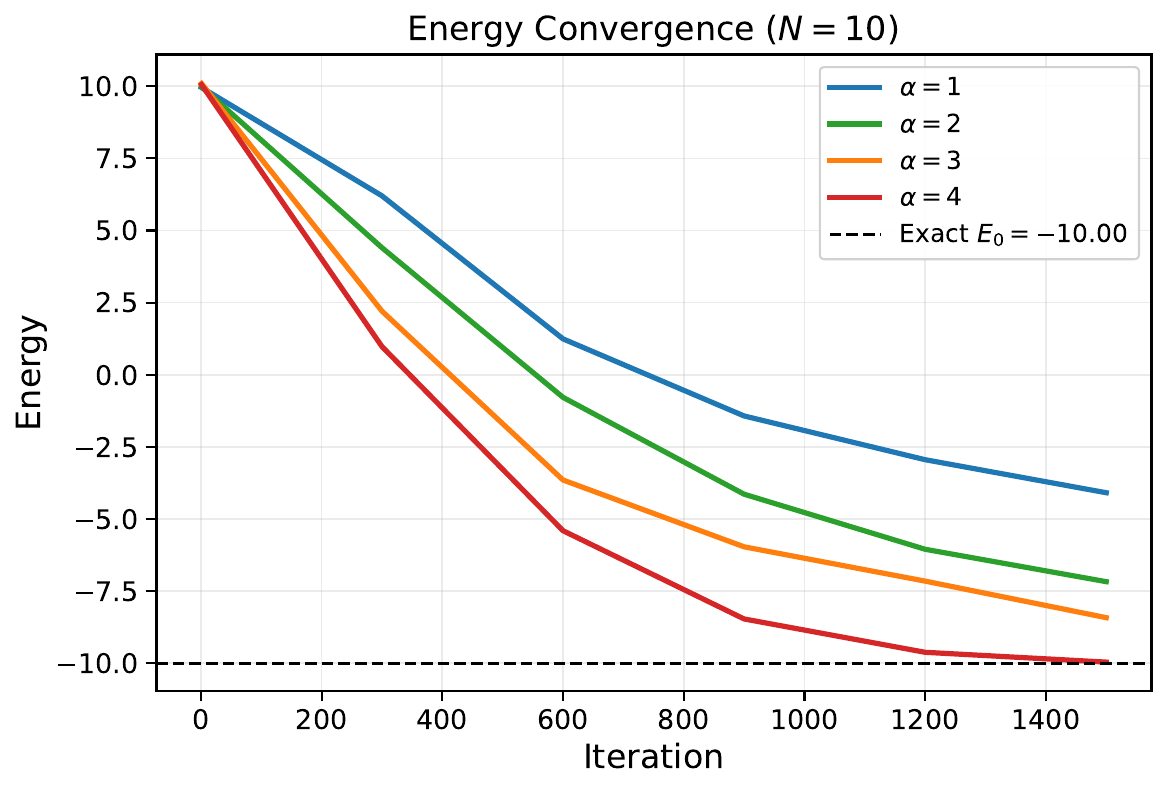}%
  \label{fig:alpha_convergence}%
}%
\hfill
\subfigure[Final energy and variance vs $\alpha$]{%
  \includegraphics[width=0.49\textwidth]{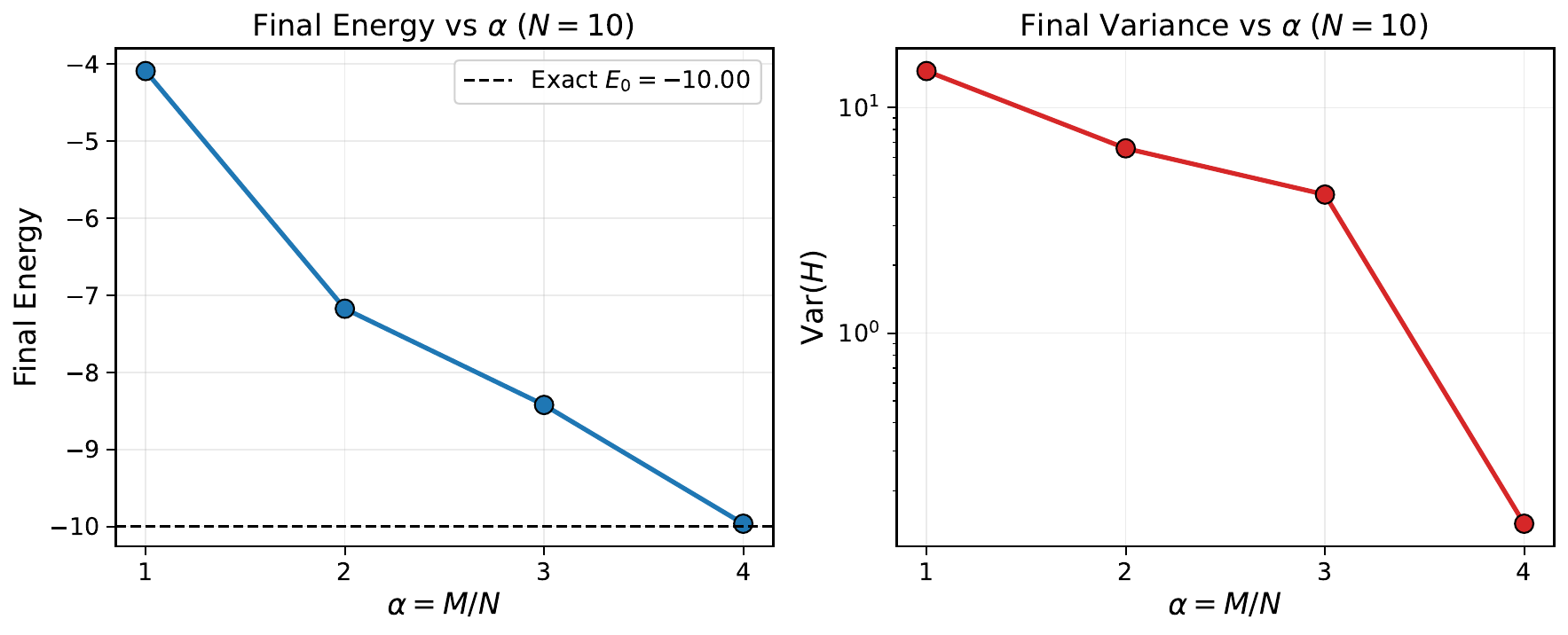}%
  \label{fig:alpha_energy}%
}
\caption{Dependence on hidden unit density $\alpha$ at fixed system size ($N = 10$). (a)~Energy convergence trajectories for $\alpha = 1, 2, 3, 4$. Higher $\alpha$ converges faster, though only $\alpha = 4$ reaches the exact ground state within 1500 iterations. (b)~Final energy and variance as a function of $\alpha$, showing monotonic improvement with increasing hidden unit density for this system size.}
\label{fig:alpha_dependence}
\end{figure}

To separate the effect of hidden unit density from system size, we fix $N$ and vary $\alpha$ (Fig.~\ref{fig:alpha_dependence}). At N=10, increasing $\alpha$ from 1 to 4 steadily improves both the final energy and variance, with $\alpha = 4$ reaching the exact ground state. The convergence rate also improves with $\alpha$, consistent with the greater representational capacity of the wider hidden layer.

The ablation analysis at $N=12$ ($\epsilon = 0.001$, 10 seeds per $\alpha$) reveals a more nuanced picture. The fraction of important units increases from $m'/M \approx 0.53$ at $\alpha = 2$ to $\sim$0.73 at $\alpha = 5$, while the energy-correlation coefficient $\rho$ peaks at $\alpha = 4$ ($\rho = 0.34 \pm 0.14$) and decreases at $\alpha = 5$ ($\rho = 0.13 \pm 0.26$). This suggests that while additional hidden units improve energy convergence, they do not necessarily improve the interpretability of the learned representation. Moderate values of $\alpha = 3$ to $4$ appear to offer a practical balance between accuracy and interpretability, consistent with prior work on one-dimensional systems \cite{carleo2017solving, borin2019physical}.

\section{Discussion}

Our analysis reveals a clear hierarchy in hidden unit importance that varies with system size. Small systems can be dominated by individual ``super units'' (like unit 3 in N=4), while larger systems distribute importance more democratically. This transition likely reflects the increasing complexity of quantum entanglement patterns in larger systems.

The phase inversion observed in N=8 correlation patterns when removing important units suggests that hidden units encode not just amplitude but also phase information critical for quantum coherence. This phase encoding may be essential for maintaining the correct quantum superposition.

The complete failure of individual units to reproduce quantum correlations, even when they strongly impact energy, demonstrates that RBMs encode quantum states through genuinely collective representations. This mirrors the quantum mechanical principle that entangled states cannot be decomposed into individual product states.

The systematic study across $N=8$ to $20$, with 279 converged models across ten seeds per configuration, adds a quantitative dimension to these observations. The fraction of important hidden units decreases with system size for all $\alpha$ values tested, with the data consistent with sublinear growth $m' \sim N^k$ with $k \approx 0.4$ across all hidden unit densities. The fit quality is moderate ($R^2 = 0.47$ to $0.78$) and the threshold sensitivity analysis confirms that the sublinear regime is robust for $\tau \in [0.15, 0.35]$. If this trend holds at larger $N$, it would imply that RBMs develop increasing redundancy as the system grows, analogous to the over-parameterization phenomenon in classical deep learning \cite{zhang2021understanding}. This raises the practical question of whether unimportant units could be pruned to reduce computational cost without degrading the variational state \cite{frankle2019lottery}.


\section{Conclusions}

We have demonstrated that RBMs trained on Heisenberg spin rings develop structured internal representations where hidden units specialize to encode antiferromagnetic quantum order. The main findings are:

\begin{enumerate}
\item \textbf{Emergent specialization and collective encoding.} Hidden units spontaneously develop antiferromagnetic weight patterns, and ablation identifies a clear subset $m'$ essential for ground-state energy and correlations. No single unit can reproduce the antiferromagnetic order alone; it emerges only through the collective action of multiple units.

\item \textbf{Size-dependent hierarchy.} Small systems (N=4) can be dominated by individual critical units, while larger systems distribute importance more broadly. Removing important units disrupts both the energy and the phase structure of correlations.

\item \textbf{Decreasing fraction of important units.} Across $N=8$ to $20$ and $\alpha = 2$ to $5$, the fraction $m'/M$ of important units decreases with system size, consistent with sublinear growth $m' \sim N^k$ ($k \approx 0.4$, $R^2$ up to $0.78$). This conclusion is robust across a range of classification thresholds $\tau \in [0.15, 0.35]$.


\item \textbf{Moderate $\alpha$ suffices.} Hidden unit densities of $\alpha = 3$ to $4$ provide the best balance between convergence quality and interpretability of the learned representation.
\end{enumerate}
\vspace{0.1em}
These results show that ablation-based interpretability, originally demonstrated on small systems, can be extended to moderate system sizes and provides physically meaningful classifications of hidden units. The observed decrease in the fraction of important units with system size is suggestive of growing redundancy, though the evidence is preliminary. Whether these patterns persist in frustrated systems, higher dimensions, or near quantum phase transitions remains to be seen.

\acknowledgments
We thank colleagues for valuable discussions. Computations were performed using NetKet \cite{netket,vicentini2022netket3}. This work was supported by Fraunhofer IAO.

\bibliography{biblio}
\end{document}